\documentclass[conference]{IEEEtran}
\usepackage[export]{adjust box}
\usepackage{pgf}
\usepackage{soul}
\usepackage{cite}
\usepackage{amsmath,amssymb,amsfonts}
\usepackage{algorithmic}
\usepackage{graphicx}
\usepackage{textcomp}
\usepackage{xcolor}
\usepackage{mathtools}
\usepackage{subfigure}
\usepackage{enumitem}
\usepackage{balance}
\usepackage{url}
\newcommand{\spara}[1]{\smallskip\noindent{\bf #1}}

\usepackage{cuted}
\def\BibTeX{{\rm B\kern-.05em{\sc i\kern-.025em b}\kern-.08em
    T\kern-.1667em\lower.7ex\hbox{E}\kern-.125emX}}
    
\begin{document}

\title{Emergency Incident Detection from Crowdsourced Waze Data using Bayesian Information Fusion
%\thanks{The work presented in this paper is supported by the National Science Foundation (NSF) under the award XXXXX.}
}

%% https://tex.stackexchange.com/questions/458204/ieeetran-document-class-how-to-align-five-authors-properly

\makeatletter
\newcommand{\linebreakand}{%
  \end{@IEEEauthorhalign}
  \hfill\mbox{}\par
  \mbox{}\hfill\begin{@IEEEauthorhalign}
}
\makeatother

\author{
\IEEEauthorblockN{Yasas Senarath}
\IEEEauthorblockA{
\textit{George Mason University}\\
Fairfax, VA, USA \\
ywijesu@gmu.edu}
\and
\IEEEauthorblockN{Saideep Nannapaneni}
\IEEEauthorblockA{
\textit{Wichita State University}\\
Wichita, KS, USA \\
saideep.nannapaneni@wichita.edu}
\and
\IEEEauthorblockN{Hemant Purohit}
\IEEEauthorblockA{
\textit{George Mason University}\\
Fairfax, VA, USA \\
hpurohit@gmu.edu}
\and 
%\linebreakand
\IEEEauthorblockN{Abhishek Dubey}
\IEEEauthorblockA{
\textit{Vanderbilt University}\\
Nashville, TN, USA \\
abhishek.dubey@vanderbilt.edu}
}

\maketitle

\begin{abstract}

The number of emergencies have increased over the years with the growth in urbanization. 
This pattern has overwhelmed the emergency services with limited resources and demands the optimization of response processes. %, resource propositioning, and allocation. 
It is partly due to traditional `reactive' approach of emergency services to collect data about incidents, where a source initiates a call to the emergency number (e.g., 911 in U.S.), delaying and limiting the potentially optimal response.  
Crowdsourcing platforms such as Waze provides an opportunity to develop a rapid, `proactive' approach to collect data about incidents through crowd-generated observational reports. 
However, the reliability of reporting sources and spatio-temporal uncertainty of the reported incidents challenge the design of such a proactive approach.  
Thus, this paper presents a novel method for emergency incident detection using noisy crowdsourced Waze data.
We propose a principled computational framework based on Bayesian theory to model the uncertainty in the \textit{reliability of crowd-generated reports} and \textit{their integration across space and time} to detect incidents.  
Extensive experiments using data collected from Waze and the official reported incidents in Nashville, Tenessee in the U.S. show our method can outperform strong baselines for both F1-score and AUC.    
The application of this work provides an extensible framework to incorporate different noisy data sources for proactive incident detection to improve and optimize emergency response operations in our communities.     

\end{abstract}


%%%%%%%%%%%%%%%%%%%%%%%%%%%%%%%%
\section{Introduction}
\label{sec:intro}

%- Why incident detection and incident rate prediction is critical for emergency management \\
Emergency response to incidents such as road accidents is one of the most pressing problems faced by communities across the globe~\cite{mukhopadhyay2020review}. Given the growing urbanization, this problem is further exaggerated and constrains the limited resources of emergency management agencies. For example, the high dependence on emergency communication lines limits and delays the ability to timely collect data about incidents, especially during disasters. At such times, 
%- Challenge: Incomplete, delayed, and inaccessible conventional observed data \\ 
%The high dependence on emergency communication lines to collect also limits and delays the ability to timely collect data about incidents. 
%
%Further, as evidenced by several recent disasters, 
the damage to centralized emergency operations centers and communication lines often make the collection of information hard and therefore crowdsourced reports, even though sporadic and possibly false, become an important tool (e.g., Hurricane Harvey in 2017
~\footnote{http://wapo.st/2iAZhbc?tid=ss\_tw\&utm\_term=.fdd1e12b125a}). 

%- Opportunity: Unconventional and rapidly accessible data from Crowdsourcing and diverse Web sources \\ 
There are a number of commercial crowdsourcing platforms. 
%The last two decades have shown an exponential growth and adoption of crowdsourcing platforms among public.
User-generated content on these platforms often provide near real-time observations of the surroundings of the source user, which can be an invaluable data source in the times of an emergency~\cite{amin2018evaluating}. However, the  inaccuracy of users (location and time of reporting) while creating observational reports about incidents 
%as an unconventional sensing mechanism for proactive emergency management
creates challenges, mainly \textit{unreliability of reporting sources} and the \textit{uncertainty of the reported incident information}
%due to the potential inaccuracies in the location and time of reporting 
~\cite{lenkei2018crowdsourced}.  

\begin{figure}[t]
\centering
    \begin{subfigure}
      \centering
      \includegraphics[width=0.45\linewidth]{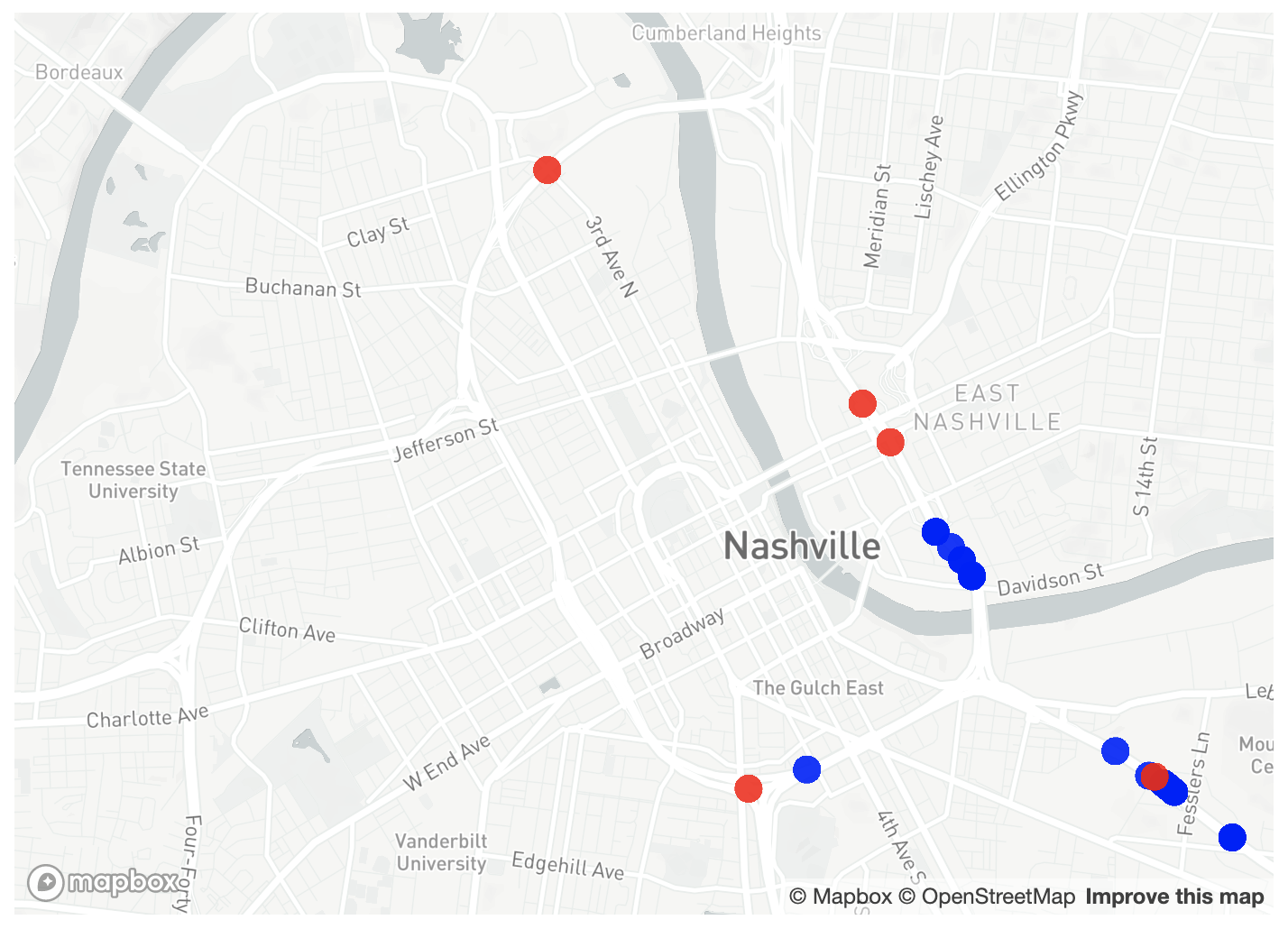}
    \end{subfigure} 
    \begin{subfigure}
      \centering
      \includegraphics[width=0.45\linewidth]{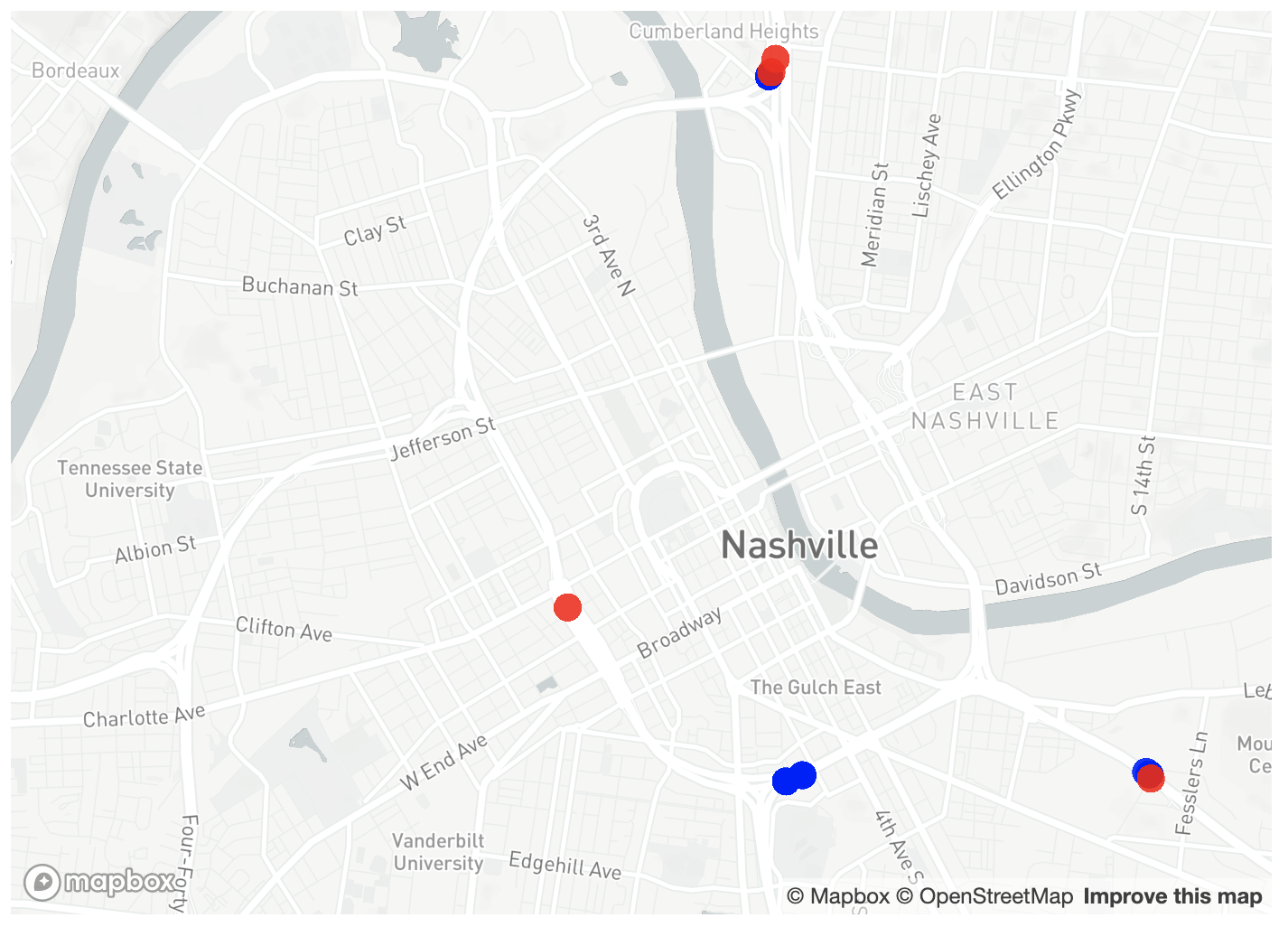}
    \end{subfigure}
%\vspace{-0.1in}
\caption{Figure shows the varied distribution of crowdsourced Waze reports (\textit{blue}) about incidents (\textit{red}) in a region at two different times, illustrating the information fusion challenge for crowdsourced data %distributed differently 
across space and time. % in same region at different time(blue: \textit{Waze}, red: \textit{E-TRIMS})
}
\label{fig:problem_illustration}
\end{figure}

%Waze provides a mobile app to users to give quicker navigation routes, while app users can report observations such as accidents, traffic jams, etc. The challenges in processing Waze data include the \textit{unreliability of reporting sources} and the \textit{uncertainty of the reported incident information}, due to the potential inaccuracies in the location and time of reporting. For instance, a source can move far away from the incident location during report, or delay reporting, %time based on convenience, the location sensor may have calibration issues, etc.    
If we consider the crowd-sourced report to be a sensor reading and the actual incident requiring emergency response to be a failure, the problem of inferring an incident from a series of sensor readings (crowd-sourced reports) is similar to the classical fault diagnostics problem in cyber-physical systems~\cite{abdelwahed2011model}, where %In system health management (physical system or software system), 
the goal is to characterize the system performance using heterogeneous sensor data, and identify any faults. In this paper, we consider the transportation network as the system of interest, and Waze crowdsourcing reports from people (human sensors) as analogous to data from mechanical sensors. 
Some of the differences between traditional diagnostics problem and the goal of incident detection analysis include: 
%\begin{enumerate}
    %\item 
    a.) Waze reports from human sensors can be unreliable when compared to mechanical sensors, 
    %\item 
    b.) The data from human sensors can be highly uncertain (e.g. location information), and 
    %\item 
    c.) There is a huge variability across the several Waze reports pertaining to the same incident due to the variability in the human sensor reporting % filing the reports 
    (\textit{c.f.} Figure~\ref{fig:problem_illustration}).
%\end{enumerate}

%We address the problem of emergency incident detection using such noisy, unconventional crowdsourced reports on Waze in Nashville 

%  for detecting conventional ETrims reported incidents and unconventional Waze
\textbf{\underline{\textit{Paper Contributions}}}: We present a novel Bayesian-theoretic method for incident detection that systematically models the uncertainty in the spatio-temporal information fusion of noisy crowdsourced reports while considering the ground-truth as conventional reports from the official local government agencies. In this paper, we analyzed Waze Reports from Nashville, Tennessee, USA and considered the incident information from the Enhanced Tennessee Roadway Information Management System (\textit{E-TRIMS}) as the ground truth.    

The proposed approach is novel as it uses a principled Bayesian-theoretic approach for aggregating uncertain crowdsourced data, and for estimating probabilities of occurrence of incidents. The Bayesian approach enables sequential updating of incident probabilities as new Waze reports arrive in time. Also, the quantitative evaluation of incident probabilities provide a measure of confidence of our prediction.

% Our experiments (Section~\ref{sec:results_detection}) show the proposed approach outperforms baselines \textcolor{red}{that uses surface features from Waze reports}.
% % \hl{we need to write a few lines on what this baseline is.}
% with absolute gain up to $3$\% in F1-score.  
% % 
% Further, we provide an in-depth analysis (Section~\ref{sec:results_st_analysis}) of the effect of space and time resolution in information fusion to detect incidents.  
% %
% Lastly, a case study (Section~\ref{sec:results_case_study}) demonstrates the benefits of leveraging the unconventional crowdsourced data in early detection of incidents than the conventional method of incident report collection.    

\textbf{\underline{\textit{Paper Organization}}}:
Section \ref{sec:related} provides a review of related work. Section \ref{sec:data} provides a review of information avaialable from incident reports from Waze and E-TRIMS platforms. Section \ref{sec:method} details the problem statement of information fusion of Waze data and describes the proposed Bayesian information fusion approach. Section \ref{sec:experiments} presents the experimental evaluation study that was carried out to demonstrate the proposed methodology. Section \ref{sec:results} presents the results of the experimental case study and a comparison of the proposed method with the existing approaches, and Section \ref{sec: conclusion} presents the concluding remarks. 
%%%%%%%%%%%%%%%%%%%
%%%%%%%%%%%%%%%%%%%
\section{Background and Related work}
\label{sec:related}

% \subsection{Crowdsourced Data in Transportation Analytics}
%
% - discuss different domain for analytics (transportation, surveillance, ..) \\ 
%Crowdsourced data (such as Twitter, Instagram, Waze, and Foursquare) represents another source of data that has been used for a variety of analyses such as emergency disaster response \cite{purohit2018structured}, urban activity detection \cite{niu2020crowdsourced}, disease transmission \cite{chunara2013we}, and traffic management \cite{amin2018evaluating}. Since the focus of this paper is on road incidents, we review previous related work on the use of crowdsourced data for it. % 
Transportation systems are the critical infrastructure of the modern cities. Therefore, significant research has been conducted to perform analytics on the transportation systems and traffic conditions using various sources of available data, including crowdsourced data. Hence, we provide a brief literature overview on the use of crowdsourced data for analyzing transportation systems.  
% Recent years have shown a growing interest in leveraging user-generated, crowdsourced data as we summarize next.
Silva et al \cite{silva2013traffic} studied the breadth and spatial coverage capabilities of the Waze report information to understand the traffic conditions, and also their limitations. They concluded that the frequency of the Waze alerts is consistent with the users' routines and traffic patterns. 
Sanchez et al. \cite{sanchez2019dynamic} investigated the impacts of dynamic traffic lighting by analyzing the traffic jam and traffic incident reports from Waze. The authors concluded that  dynamic traffic lighting can reduce traffic congestion and improve traffic speed but has no impact on the occurrence of traffic accidents. 
%
%Santos et al. \cite{dos2017analyzing} also performed a comparative study between the official accident (BHTrans) and Waze datasets in the city of Belo Horizonte, Brazil, and found that there was a 7\% match in the accidents reported by both the datasets. When compared to Waze reports, the BHTrans data was spread more uniformly across the day. Most of the accidents reported by Waze but not by BHTrans were classified as having lower severity. Accidents reported only by BHTrans were concentrated on the central region of Belo Horizonte, while the ones reported by Waze were mostly on highway segments. 
% 
Eriksson \cite{eriksson2019towards} and Santos et al. \cite{dos2017analyzing} integrated accident data from the official accident database and Waze to create a comprehensive accident database after removing the data that were common in both the datasets through spatio-temporal proximity analysis. 

Lenkei \cite{lenkei2018crowdsourced} performed a comparative study of the accidents reported through Waze and the official traffic database in Sweden (Trafikverket), and concluded that there was a 43\% overlap in the accident data between Waze and Trafikverket. 
The author provided two reasons for the increased number of Waze alerts. First, some of the Waze alerts can either be self-resolving, false alarms, or of low impact. And, second, the Waze data stream can include multiple alerts referring to the same accident. %The author built a binomial logistic regression model using a set of independent variables (type of alert, the type of the road where the incident happened, the part of the day when the incident happened, the rank of the reporter, the number of related alerts, the number of related jams and the number of confirmations that the alert has received) to determine if a Waze alert refers to an incident that should be handled by incident management authorities. 
Lenkei also observed that 27.5\% of the incidents in Trafikverket were detected earlier by Waze. 

These studies motivate our work to explore the predictability of unconventional, \textit{noisy} Waze reports to detect emergency incidents, and especially, the need for a generic, principled approach for spatio-temporal information fusion of the \textit{redundant} and \textit{noisy} crowdsourced reports to detect emergency incidents.  

Furthermore, Flynn et al. \cite{flynn2018estimating} investigated the ability of Waze data to serve as a reliable indicator of police-reportable crashes through machine learning approaches. In particular, the authors considered a Random Forest (RF) model to predict a police-reportable crash (binary outcome) using a set of  %independent variables
features based on weather, Waze Reports (median report reliability, number of records, type of Waze event, road classification), time (hour of day and day of week), and urban area classification. 

In summary, we note that while there exist prior research studies on crowdsourced data for event detection and transportation incidents, there is a lack of
%exploration for
a.) a principled approach for information fusion of noisy data to detect incidents that could be extended in general across crowdsourcing platforms because the existing approaches consider only simple aggregation methods (count/average) of surface level features without considering impacts of uncertainty associated with report integration, and 
b.) a comprehensive analysis of the effect of spatio-temporal resolution in the fusion process of noisy crowdsourced data.

Next, we present our methodology to remedy these gaps in the prior research. First, we provide a brief overview of the Waze and E-TRIMS datasets (with similar trends, \textit{c.f.} Figure~\ref{fig:waze_tdot_hourly_dist}) used in this paper to train a model for incident prediction.     

\begin{figure}[ht!]
    \vspace{-0.15in}
    \centering
    \includegraphics[scale=0.45]{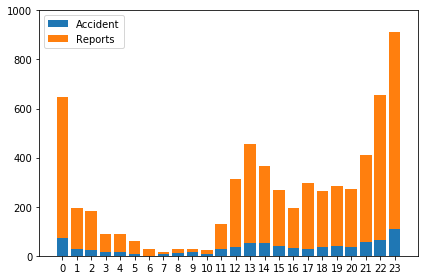}
    \vspace{-0.15in}
    \caption{Similar hourly trend of Waze reporting and accident records in the  official E-TRIMS system.} 
    \vspace{-0.15in}
    \label{fig:waze_tdot_hourly_dist}
\end{figure}

%%%%%%%%%%%%%%%%%%%%
%%%%%%%%%%%%%%%%%%%%
\begin{figure*}[ht!]
    \centering
    \includegraphics[scale=0.3]{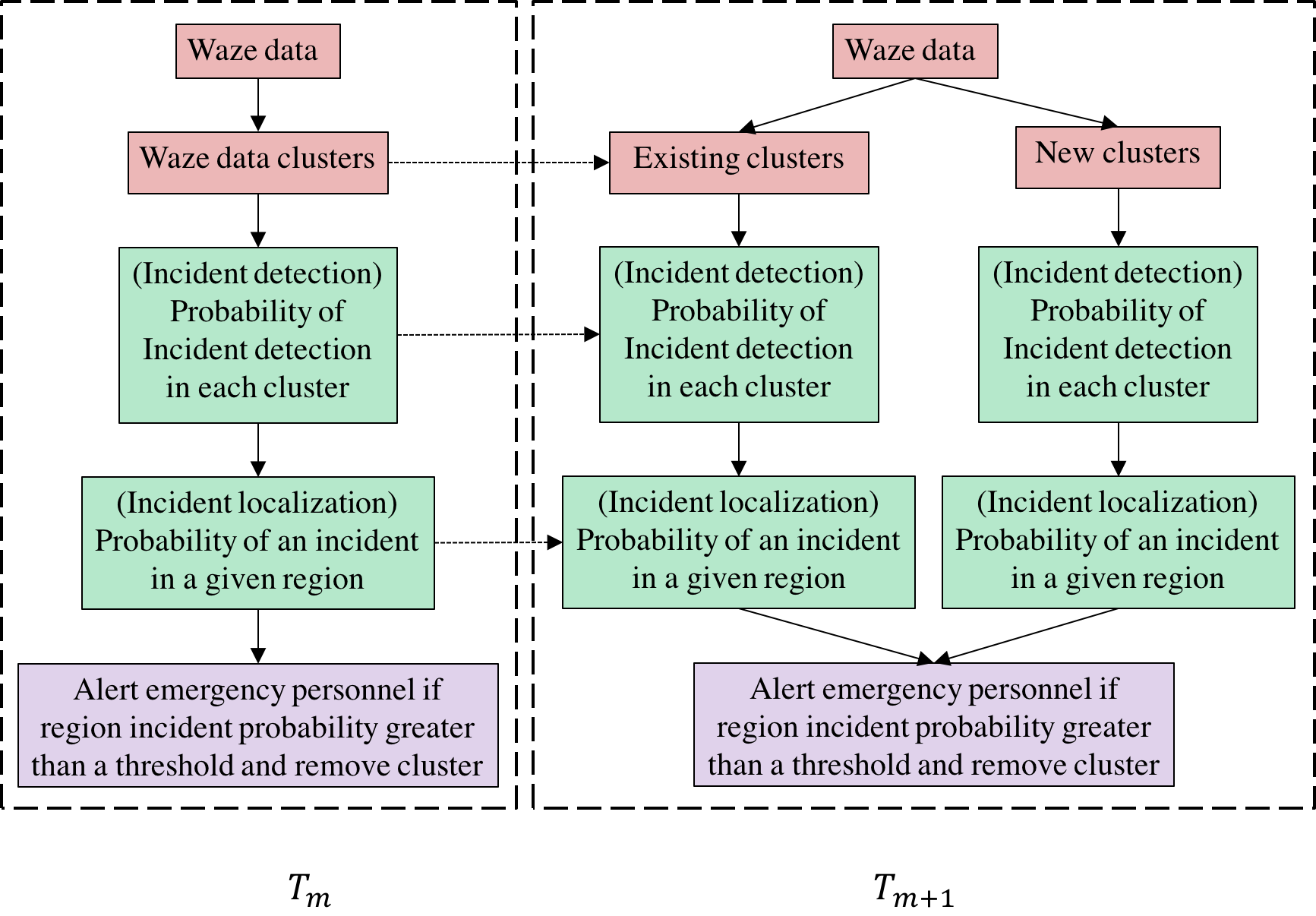}
    \caption{Overview of the Bayesian-theoretic approach to fuse information and detect incidents.} 
    \label{fig:overview}
    \vspace{-0.1in}
\end{figure*}

\section{Data platforms}
\label{sec:data}
% We first describe the data platforms in this study, followed by definitions and notations to formally state our problem and then, present our methodology. 

%%%%%%%%
\subsection{Waze Platform} \label{subsec:waze_dataset}
Waze is a GPS navigation application that enables its users to send reports of different traffic conditions while they are travelling~\cite{waze_driving_2020}. In the following, we provide the descriptions of most relevant attributes of a user-generated report: % that are relevant in solving our problem.
\begin{itemize}[leftmargin=*]
    \item \textbf{type}: Type of traffic alert; this study only uses traffic alerts with type `ACCIDENT'.
    \item \textbf{confidence}: a score for the report based on other users' reactions to the report, between [0-10]. %created by the reporting user.
    \item \textbf{reportRating}: The level of the reporting user credibility, between [1-6].
    \item \textbf{reliability}: Indicates the reliability of the Waze report, where the reliability is based on \textit{confidence} of the report and \textit{reportRating} score of the user; value between [1-10]. 
    \item \textbf{location}: Geographic coordinates (latitude, longitude) of the report origin. 
    \item \textbf{pubMillis}: Reporting time in milliseconds since epoch. 
\end{itemize}

%%%%%%%%%
\subsection{E-TRIMS Platform}

%We use data from 
%Enhanced Tennessee Roadway Information Management System (
This is a single integrated system in Tennessee state of the U.S. that includes State and local roadways, structures, pavement, traffic, photo logs, and crash data. We use this data source to validate the models for incident detection. In contrast to crowdsourced data from Waze, the E-TRIMS traffic data is more reliable. %\textcolor{red}{and curated with the help of official emergency services' databases}. 
An incident record in the E-TRIMS constitutes of attributes that are similar to user-generated report on Waze: 

\begin{itemize}[leftmargin=*]
    \item \textbf{latitude}: GPS Coordinate Latitude of the E-TRIMS incident record.
    \item \textbf{longitude}: Longitude of the E-TRIMS incident record.
    \item \textbf{timestamp}: Time the accident is recorded in the system in milliseconds since epoch. 
    \item \textbf{unit\_segment\_id}: Id of the road segment that accident has happened. 
\end{itemize}

\section{Problem Statement \& Proposed Methodology}
\label{sec:method}

\begin{table}[h]
\caption{Notation for Variables in Methodology} 
    \label{tab:notation}
    \centering
    \begin{tabular}{|p{1cm}|p{6.5cm}|}
    \hline
    \textbf{Variable} & \textbf{Description} \\
    \hline
    $W$ & Total area of interest (e.g., City of Nashville)\\
    \hline
    $k$ & Number of non-overlapping regions $W$ is divided into\\
    \hline
    $R_j$ & $j^{th}$ region\\
    \hline
    $t_s$ & time interval after which Waze reports are aggregated\\
    \hline
    $T_m$ & $m^{th}$ time step\\
    \hline
    $\mathbf{w}_m$ & collection of Waze reports available in $T_m$\\
    \hline
    $h$ & Number of clusters that the Waze reports are divided into\\
    \hline
    $C_u$ & $u^{th}$ cluster\\
    \hline
    $\mathbf{w}_{mu}$ & collection of Waze reports available in $T_m$ in $C_u$\\
    \hline
    $n$ & Number of Waze reports in $\mathbf{w}_{mu}$\\
    \hline
    $w_{mu}^i$ & $i^{th}$ Waze report in $\mathbf{w}_{mu}$\\
    \hline
    $l_{mu}^i$ & latitude value associated with $w_{mu}^i$\\
    \hline
     $g_{mu}^i$ & longitude value associated with $w_{mu}^i$\\
    \hline
     $x_{mu}^i$ & latitude and longitude information associated with $w_{mu}^i$, i.e., $x_{mu}^i = [l_{mu}^i,g_{mu}^i]$\\
    \hline
    $R_{ju}$ & $j^{th}$ region covered by $C_u$, i.e., the $u^{th}$ cluster\\
    \hline
    $v_u$ & Number of regions in the area covered by the $u^{th}$ cluster, $C_u$\\
    \hline
    $r$ & reliability score associated with a Waze report\\
    \hline
    $p$ & probability value associated with the reliability score of a Waze report\\
    \hline
    $I_u$ & binary variable that relates to the occurrence of an incident in the $u^{th}$ cluster. $I_u=1$ and $I_u=0$ relates to the occurrence and absence of an incident respectively in the $u^{th}$ cluster\\
    \hline
    $P(.)$ & Probability function\\
    \hline
    $r_{mu}^i$ & Reliability associated with $w_{mu}^i$\\
    \hline
    $p_{mu}^i$ & Probability associated with the reliability of $w_{mu}^i$\\
    \hline
    $t_r$ & Reaction time of a reporter between noticing an incident and reporting it on Waze\\
    \hline
    $v_t$ & Average velocity of a Waze reporter\\
    \hline
    $\delta$ & Distance traveled by a Waze reporter in the duration $t_r$\\
    \hline
    $A(w_{mu}^i)$ & Area around the location of a Waze report, i.e., $x_{mu}^i$ in which an incident may have happened\\
    \hline
    $\Omega(m,n)$ & Function that computes area overlap between any two areas $m$ and $n$\\
    \hline
    
    \end{tabular}
\end{table}

%%%%%%%
\subsection{Definitions and Problem Statement} \label{subsec:def}
We summarize the notations used in Table~\ref{tab:notation} and define the key concepts as follows: 
\begin{itemize}[leftmargin=*]
    \item \textbf{Region ($R_j$)}: We divide the area of interest ($W$) for incident detection into hexagonal grids called regions, similar to the approach used in \cite{mukhopadhyay2019online}. % has used a similar grid based approach for responder dispatch. 
    We use H3: Uber’s Hexagonal Hierarchical Spatial Index~\cite{uber_h3_2020}. This enables us to choose a resolution value ($Res$) and obtain grid indexes corresponding to provided geo-coordinates. Regions are the spatial unit of accident prediction in this study.
    \item \textbf{Incident Time Period ($T'$)}: We assume that an incident in a region $R_j$ will prevail for a $T^\prime$ period. We call this period an incident time period in $R_j$ and assume a single incident occurring in $R_j$ during $T^\prime$.  
    \item \textbf{Time step ($T_m$)}: The smallest time interval at which Waze reports are aggregated ($T_m \in T^\prime$) to detect an incident. %This is the temporal unit of incident detection in this study. 
    %\item \textcolor{red}{\textbf{anything else} as definition?} .. 
\end{itemize}

\textbf{\underline{Motivating Example:}} Let us consider a  scenario to understand the problem that is considered in this paper.

Assume that we begin collecting Waze reports at 12:00 am in the intervals of five minutes. Let there be six Waze reports in the first time step (five minutes) and four Waze reports in the second time step, and similarly, we may have one or more Waze reports in each time step. We would like to use these Waze reports to detect the existence of one or more traffic incidents, and also predict the location of the incident(s).

Given the six Waze reports in the first time step, we will need to infer if these Waze reports are associated with the same traffic incident or multiple incidents. Then, we will need to fuse multiple Waze reports associate with same incident to predict if a traffic incident did occur and also its location. 

In the next time step, we have four Waze reports. Now, we will need to decide if these Waze reports are associated with the same incidents identified in the previous time step or if they correspond to new incident(s) not observed in the previous time step. After associating Waze reports to incidents, we will predict the presence of an incident and its location. The subsequent question is to decide when to alert the emergency response authorities about a traffic incident.

\textbf{\underline{Problem Statement}}: Given a time series of crowdsourced Waze reports, we will need to answer the following questions:
\begin{enumerate}
    \item Associate available Waze reports to one or more incidents (Sections IV-C and IV-G)
    \item Aggregate the Waze reports associated with an incident to estimate the probability of that incident (Section IV-E)
    \item Predict the locations of incidents (Section IV-F)
    \item Determine when to alert the emergency response authorities about potential incidents (Section IV-H)
\end{enumerate}

\subsection{Overview of the proposed approach} 
\label{subsec:overview}

Figure \ref{fig:overview} presents an overview of our proposed approach. Since Waze reports are available on  continuum in time and space (e.g., city of interest), we implement spatio-temporal discretization and analyze the Waze reports to reduce computational expense and facilitate real-time analysis. After spatio-temporal discretization, we will implement a clustering (grouping) approach on the Waze reports to create Waze data clusters such that all the Waze reports in a cluster are talking about the same incident. The ``red" boxes correspond to data gathering and clustering.

We then aggregate the incident information from various Waze reports using a Bayesian information fusion approach to calculate the probability of incident and also calculate the probability of the incident in a given region. The ``green" boxes correspond to the %calculation of 
incident detection and localization.

If the incident probability in a region was found to be greater than a threshold value, we alert the local emergency response personnel. The threshold value is learned using the Waze reports and incident information available from the local law enforcement agencies. The ``purple" box represents decision-making. The threshold value is learned from data by using the E-TRIMS dataset as the ground truth dataset.

If the probability of an incident in any region is less than the threshold, we then consider Waze reports in the next available discrete time step to update the probability of an incident in a region. In this way, we perform sequential updating of incident probabilities. We will also consider a pre-determined number of time steps for updating the incident probabilities. If the incident probability is less than the threshold at the end of all the time steps, we assume an incident did not happen and ``delete" the cluster.  We explain each of the analysis steps with additional details below.

% \hl{again the overview should provide some example and make it easier to understand. Right now without the figure 5 this is impossible to understand. please bring figure 5 here and then explain the whole section and overview using the figure. -added image here}

%\textcolor{red}{Saideep: it's better to provide an overview of our solution to the above problem before we give details in subsections.} \textcolor{blue}{DONE}

%%%%%%%%%%%%%%%%%%%%
\subsection{Spatio-Temporal Discretization and Grouping}
\label{subsec:grouping}

% \textcolor{red}{Here, we need to talk about time step and time interval}
We perform %clustering analysis to
grouping of several Waze reports into discrete groups where all the reports in a given cluster are assumed to be associated with the same incident. Let $T_m$ represent the $m^{th}$ time step; therefore, $T_m - T_{m-1} = t_s$. Let $\mathbf{w}_m$ represent the set of reports that are available at $T_m$. Let $C_u, u=1 \dots h$ reprsent the $h$ clusters to which $\mathbf{w}_m$ are grouped into. Let $\mathbf{w}_{mu}$ represent the set of $n$ reports in a cluster $C_u$. 
We explore two strategies to group Waze reports that identify same incident: segmentation and density-based clustering.

\textbf{Segmentation Approach}: In segmentation method, at the start of incident time period $T^\prime$, we begin to collect Waze reports when the first report appears and stop collecting until %end of %reports for that incident once 
$T^\prime$ is over. %Any Waze report that appears after the $T^\prime$ period is considered for a new incident. 
%\textcolor{red}{ 
The process is repeated for each region $R_j$. We consider all the surrounding regions ($R_{ju}$) that intersects Waze reports as the region covered by this grouping approach. 
%} 
Advantages of this approach %over clustering approach are 
include simplicity and ease of identifying the segments. 
% 
%Since Waze reports can arrive any time, it would be computationally expensive and sometimes not feasible to perform incident detection after every Waze report. Therefore, we perform the analysis at discrete time steps. Let $t_s$ represent the length of each time step at which the Waze reports are aggregated. After every time step, we may have several Waze reports scattered across several regions, which maybe associated with one or more incidents. \textcolor{blue}{Yasas: This paragraph is talking about spatio-temporal discretization. We can combine this with your writeup}

\textbf{Clustering Approach}: In addition to following the segmentation process, this approach uses DBSCAN~\cite{ester1996density} to identify density-based clusters of Waze reports by providing location and time information as features. We explore different values of $\epsilon$ parameter in DBSCAN ranging from 0.5 to 0.9, incriminating at 0.1, to find the best set of clusters. For $\epsilon = 0.8$, we obtained the best silhouette score of $0.8156$. The resulting clusters are used in our approach to detect incidents.  Figure~\ref{fig:cluster_dist} provide the distribution of the cluster time period calculated using Equation~\ref{eqn:cluster_time_period}, where $\mathbf{w}_{u}$ indicates Waze reports belonging to a cluster $C_u$. Clusters with zero $Time Period$ are excluded in the figures for clarity.

%\textcolor{red}{
Hereafter the term cluster ($C_u$) is used to denote the group of Waze reports belonging to the same incident independent of the grouping method employed, unless explicitly mentioned.
%}

\vspace{-0.1in}
\begin{equation}
\label{eqn:cluster_time_period}
\begin{aligned}
    Time Period = max(\mathbf{w}_{u}.pubMillis) - min(\mathbf{w}_{u}.pubMillis)
\end{aligned}
\end{equation}

\begin{figure}[ht!]
    \centering
      \includegraphics[scale=0.6]{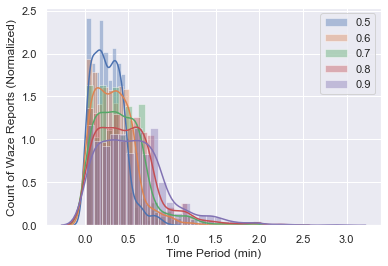}
      \caption{Distribution of time interval (min) of clusters. Colors indicate the value of $\epsilon$.} 
      \label{fig:cluster_dist}
      \vspace{-0.1in}
\end{figure}

% \begin{figure}[ht!]
%     \centering
%       \includegraphics[scale=0.45]{figs/clustering/cluster_dist_spec.png}
%       \caption{Distribution of time interval (min) of clusters that resulted with maximum silhouette score (when $\epsilon = 0.8$).} 
%       \label{fig:cluster_dist_spec}
% \end{figure}

% \textcolor{red}{Yasas: plz add the figure from slides for clustering results}
% \begin{figure*}[ht!]
%     \centering
%     \includegraphics[scale=0.40]{figs/method_summary.png}
%     \caption{Overview of the Bayesian-theoretic approach to fuse information and detect incidents} 
%     \label{fig:overview}
% \end{figure*}
%%%%%%%%%%%%%%%%%%%%
\subsection{Data Preprocessing}
\label{sec:preprocessing}

In data preprocessing stage, We calculate prior probability with E-TRIMS data for each hour of the day of each region. Equation~\ref{eqn:prior_calculation} illustrates how we calculate the priors for region $R_j$ at hour $H$ of the day. Function $COUNT(region, hour)$ returns the count of E-TRIMS accident records belonging to the provided region and hour of the day from a predefined set of E-TRIMS accident records. $COUNT(*)$ is the number of records in predefined set of E-TRIMS accident records.

\begin{equation}
\label{eqn:prior_calculation}
\begin{aligned}
    P(I_u=1|R_j, H) = \frac{COUNT(R_j, H)}{COUNT(*)}
\end{aligned}
\end{equation}

%%%%%%%%%%%%%%%%%%%%
\subsection{Incident Detection}
\label{subsec:pred}

% Our approach for incident detection and localization is similar to the approaches commonly used in fault detection and localization analyses performed in system health management. 
% MOVED. 
%\textcolor{blue}{Hemant: I think we can move the above information to the introduction section. It is good to mention that this problem is similar to fault diagnostics problem at the beginning of the paper}

Let $W$ represent the total area of interest for incident detection, and let $W$ be divided into $k$ non-overlapping regions, $R_j, j=1\dots k$. Hence, $R_j \cap R_k = \emptyset, j\neq k$ and  $\cup_{j=1}^k R_j = W$. These regions can be of different sizes depending on the analysis requirements.

%To this end, 
%\textcolor{blue}{Yasas: This paragraph is talking about clustering. Since you mentioned this earlier, you can combine this paragraph with your writeup}

As discussed in Section \ref{subsec:waze_dataset}, each Waze report is associated with location information, i.e., latitude and longitude values. Let $w_{mu}^i$ represent the $i^{th}$ Waze report in the $p^{th}$ cluster at $T_m$. Let $x_{mu}^i = [l_{mu}^i,g_{mu}^i]$ represent the location of $w_{mu}^i$.

% Within each cluster, we calculate the mean location using the location information of all the Waze reports in the cluster, i.e., $x_{mu}^C = \bigg[\frac{\sum_{I_u=1}^n l_{mu}^i}{n},\frac{\sum_{I_u=1}^n g_{mu}^i}{n}   \bigg]$.

% With $x_{mu}^C$ as the center, we consider a circle of radius $s_p$\textcolor{red}{(Need to explain how the radius is calculated)}, and we assume that an incident occurred in one of the regions that has an overlap with the circle. 

Let $R_{ju}, j=1\dots v$ represent the $v$ regions that are covered by the Waze reports, and we assume that an incident may have occurred in any of these $v$ regions. Using available Waze reports, we calculate the probability of an incident in each region. Under the assumption that an incident occurred, this probability represents our belief of that incident in each region. 

%Each Waze report is associated with a reliability score. 
%The reliability score is an integer value between 1 and 10. In order to calculate incident detection, we need to aggregate multiple Waze reports with different reliability scores. Since reliability scores are empirical scores, we do not have a mathematical framework to aggregate them. We also know that as the reliability increases, the overall confidence regarding the report incident also increases. Therefore, we associated every possible reliability score with a probability value. 
If $r$ is the reliability score ($ r \in \mathbb{Z}, 1\leq r \leq 10$), then the associated probability measure is denoted as $p$ ($0\leq p \leq 1$). 
%We will have ten different probability values corresponding to the ten reliability values. Moreover if $r_i$ and $r_j$ are two different reliability scores such that $r_i < r_j$, then the associated probability values $p_i$ and $p_j$ also follow that $p_i < p_j$. %\textcolor{red}{Accordingly, 
In our current approach, we divide $r$ of each Waze report by ten (max) to obtain the probability $p$, future work can explore different mechanism to estimate it. 
%since it is simple and satisfies the aforementioned conditions.} \textcolor{blue}{Yasas: the above paragraph is a general description of Waze report. We can move it to the background section}

% \textcolor{red}{Here, we need to discuss how to calculate $p_i$ and $p_j$ from associated reliability values.}

% \hl{This whole section is very hard to read -- saideep please review and fix this.}
% The problem that we are solving in this step can be described as follows.

% \textit{What is the likelihood of the occurrence of an incident given a set of Waze reports? This can be answered by generating a incident (analogous to a fault) hypothesis that the availble Waze reports do correspond to an incident and then test, the hypothesis testing using Bayesian analysis.}

Let $I_u=1$ and $I_u=0$ represent the event of an incident happening in any of the $v_u$ regions, and the event of no incident respectively. Let $P(I_u=1)$ and $P(I_u=0)$ represent the prior probabilities of an incident and no incident respectively. Let $P(I_u=1, R_{ju})$ represent the prior probability of an incident in region $R_{ju}$; therefore,  $P(I_u=1) = \sum_{j=1}^{v_u} P(I_u=1, R_{ju})$.

Our goal is to obtain the posterior probability of an incident happening in $R_{ju}$ using the available Waze reports $\mathbf{w}_{mu}$; the posterior probability is denoted as $P(I_u=1, R_{ju}|\mathbf{w}_{mu})$, which can be decomposed into a product of two terms as $P(I_u=1, R_{ju}|\mathbf{w}_{mu}) = P(I_u=1|\mathbf{w}_{mu}) \times P(R_{ju}|I_u=1, \mathbf{w}_{mu})$. The first term, $P(I_u=1|\mathbf{w}_{mu})$, represents the total probability that an incident happened given a set of Waze reports. Given an incident occurred, the second term, $P(R_{ju}|I_u=1, \mathbf{w}_{mu})$ computes the probability that it occurred in region $R_{ju}$.

The probability that an incident occurred or did not occur is calculated using a Naive Bayes approach. For that, we have two classes: (1) Incident occurred ($I_u=1$), and (2) Incident did not occur ($I_u=0$). Using Bayes theorem, the probability of an incident occurrence can be calculated as

\begin{equation}
\label{eqn:Incident_Bayes}
\begin{aligned}
     P(I_u=1|\mathbf{w}_{mu})& =  \dfrac{P(\mathbf{w}_{mu}|I_u=1) P(I_u=1)}{P(\mathbf{w}_{mu})}\\
    %  & \propto P(\mathbf{w}_{mu}|I_u=1) P(I_u=1) \\
     & \propto \prod_{i=1}^n P(w^i_{mu}|I_u=1) P(I_u=1)
\end{aligned}
\end{equation}

In Eq. \ref{eqn:Incident_Bayes}, $P(\mathbf{w}_{mu}|I_u=1)$ represents the likelihood of observing $\mathbf{w}_{mu}$ given that an incident occurred. Assuming all the $n$ Waze reports are independent to each other, the joint likelihood of observing all the $n$ Waze reports can be calculated as the product of the likelihood of the observing individual Waze reports. Similarly, the posterior probability of the absence of an incident can be computed as $P(I_u=0|\mathbf{w}_{mu}) \propto \prod_{i=1}^n P(w^i_{mu}|I_u=0) P(I_u=0)$. Therefore, the posterior probability of the occurrence of an incident can be calculated through the normalization, as sum of the two probabilities $P(I_u=1|\mathbf{w}_{mu})$ and $P(I_u=0|\mathbf{w}_{mu})$ should add to unity. Therefore, 

\vspace{-0.18in}
\begin{equation}
\label{eqn:post_incident}
    P(I_u=1|\mathbf{w}_{mu}) = \dfrac{\prod_{i=1}^n P(w^i_{mu}|I_u=1) P(I_u=1)}{\sum_{z=0,1} \prod_{i=1}^n P(w^i_{mu}|I_u=z) P(I_u=z)}
\end{equation}

Let $p_{mu}^i$ represents the probability associated with the reliability $r_{mu}^i$ of a Waze report, then the probability that a Waze report corresponds to a true incident is equal to $p_{mu}^i$ (sometimes referred to as true positive probability). Hence, $1-p_{mu}^i$ can be interpreted as the false positive probability, i.e., probability of a Waze report in the absence of an incident. Eq. \ref{eqn:post_incident} can be written as

\vspace{-0.18in}
\begin{equation}
\label{eqn:post_incident2}
\begin{aligned}
     &P(I_u=1|\mathbf{w}_{mu}) \\
     &= \dfrac{\prod_{i=1}^n p^i_{mu} P(I_u=1)}{\prod_{i=1}^n p^i_{mu} P(I_u=1) + \prod_{i=1}^n (1-p^i_{mu}) P(I_u=0)}   
\end{aligned}
\end{equation}

\subsection{Incident localization}
\label{subsec:local}

% The problem that we solving in this step can be written as follows:

% \textit{Given that an incident occurred, which of the regions covered by the cluster is the most likely region of the incident? In order to answer this question, we will generate an incident (analogous to a fault) hypothesis that an incident (if occurred) would have occurred in a given region and test that hypothesis with the available Waze reports.} The analysis is described below.

Typically, after a Wazer identifies an incident, it will take some time for the Wazer to report the incident. First, the user needs to process the incident, decide whether to file a Waze report, and then file a Waze report. The time it takes to perform the above three steps can vary between people; therefore, the location information does not correspond to the true location of the incident, and the incident region is in ``close" vicinity to the location in a Waze report. If $l_{mu}^i$ and $g_{mu}^i$ are the latitude and longitude values corresponding to $w_{mu}^i$, then we consider an area covered by a circle with center $(l_{mu}^i, g_{mu}^i)$ and radius $\delta$. Assume that $t_r$ is the average reaction time of a typical Wazer. 
%(\textcolor{red}{If we have some time, then we can look into some driver reaction time studies.}). 
If $v_t$ is the velocity of the Wazer, then the distance travelled by a Wazer is $v_tt_r$.
We acknowledge that there exists a slight variation of speed limits on various roads/regions, and also variation in the velocities across individual Wazers. In this paper, we do not consider such variations, and assume a constant speed. Therefore, $\delta = v_tt_r$, and is a fixed value.

Let $x_{mu}^i = (l_{mu}^i, g_{mu}^i)$ denote the location of a $w_{mu}^i$, and $\mathbf{x}_{mu}$ represents the locations of all Waze reports $\mathbf{w}_{mu}$. The probability of observing $\mathbf{w}_{mu}$ at $\mathbf{x}_{mu}$ given the occurrence of an incident at a region $R_{ju}$ can be calculated as 

\vspace{-0.18in}
\begin{equation}
\label{eqn:post_region}
\begin{aligned}
    P(R_{ju}&|I_u=1, \mathbf{w}_{mu}) \\
    & = \dfrac{P(\mathbf{w}_{mu}|I_u=1, R_{ju}) P(R_{ju}|I_u=1)}{P(\mathbf{w}_{mu}|I_u=1)}\\
    & \propto P(\mathbf{w}_{mu}|I_u=1, R_{ju}) P(R_{ju}|I_u=1)\\
    & \propto P(\mathbf{w}_{mu}|I_u=1, R_{ju}) P(R_{ju}, I_u=1)\\
%    & \propto \prod_{i=1}^n P(w_m^i|I_u=1, R_{ju}) P(R_{ju}, I_u=1)
\end{aligned}
%\vspace{-0.03in}
\end{equation}

In Eq. \ref{eqn:post_region}, $P(\mathbf{w}_{mu}|I_u=1, R_{ju})$ is the likelihood of observing $\mathbf{w}_{mu}$ at $\mathbf{x}_{mu}$ given that an incident occurred in $R_{ju}$. $P(R_{ju}|I_u=1)$ is the prior conditional probability of an occurrence of an incident in $R_{ju}$ given an incident occurred. As $P(R_{ju}|I_u=1) = \dfrac{P(R_{ju},I_u=1)}{P(I_u=1)}$, $P(R_{ju}|I_u=1) \propto P(R_{ju},I_u=1)$. $P(R_{ju}|I_u=1)$ is the prior probability of an incident occurrence in $R_{ju}$ before observing reports $\mathbf{w}_{mu}$. 

Assuming independence between the reports, $P(\mathbf{w}_{mu}|I_u=1, R_{ju})$ is equal to $\prod_{i=1}^n P(w_{mu}^i|I_u=1, R_{ju})$. The probability that a report $w_{mu}^i$ represents an incident in region $R_{ju}$ can be the fraction area overlap between the area covered by $w_{mu}^i$ and $R_{ju}$. If $A(w_{mu}^i)$ represents the area covered by $w_{mu}^i$, and $\Omega(m,n)$ the overlap function between any two areas $m$ and $n$, then the likelihood, $P(w^i_{mu}|I_u=1, R_{ju})$, is proportional to the area of overlap between the region covered by $w_{mu}^i$ and $R_{ju}$ written as $P(w^i_{mu}|I_u=1, R_{ju}) \propto \Omega(A(w^i_{mu}), R_{ju})$. Thus, Eq. \ref{eqn:post_region} can be written as

\vspace{-0.18in}
\begin{equation}
\label{eqn:post_region_area}
\vspace{-0.05in}
\begin{aligned}
    P(R_{ju}&|I_u=1, \mathbf{w}_{mu})  \\
    & \propto P(\mathbf{w}_{mu}|I_u=1, R_{ju}) P(R_{ju}, I_u=1)\\
    & \propto \prod_{i=1}^n P(w_{mu}^i|I_u=1, R_{ju}) P(R_{ju}, I_u=1)\\
    & \propto \prod_{i=1}^n \Omega(A(w^i_{mu}), R_{ju}) P(R_{ju}, I_u=1)
\end{aligned}
\end{equation}

Given $I_u=1$, an incident must occur in one of the $v$ regions, $R_{yu}, y=1\dots v$; therefore, $\sum_{y=1}^v P(R_{yu}|I_u=1, \mathbf{w}_{mu}) = 1$. Through normalization, the posterior incident probability given the occurrence of an incident is: %can be calculated as 

\vspace{-0.18in}
\begin{equation}
\label{eqn:post_region_normalization}
\begin{aligned}
    & P(R_{ju}|I_u=1, \mathbf{w}_{mu})\\
    & = \dfrac{\prod_{i=1}^n \Omega(A(w^i_{mu}), R_{ju}) P(R_{ju}, I_u=1)}{\sum_{y=1}^{v_u} \bigg(\prod_{i=1}^n \Omega(A(w^i_{mu}), R_{yu}) P(R_{yu}, I_u=1)\bigg)}
\end{aligned}
\end{equation}

The posterior probability of an incident occurring in $R_{ju}$, i.e., $P(I_u=1, R_{ju}|\mathbf{w}_{mu})$ can be calculated as a product of $P(I_u=1|\mathbf{w}_{mu})$ (Eq. \ref{eqn:post_incident2})  and $P(R_{ju}|I_u=1, \mathbf{w}_{mu})$ (Eq. \ref{eqn:post_region_normalization}).  
% \bigskip
% \begin{strip}
% \begin{align}
% \label{eqn:posterior_exact}
% \begin{aligned}
%       P(I_u=1, R_{ju}|\mathbf{w}_{mu}) = \dfrac{\prod_{i=1}^n p^i_{mu} P(I_u=1)}{\prod_{i=1}^n p^i_{mu} P(I_u=1) + \prod_{i=1}^n (1-p^i_{mu}) P(I_u=0)} \times \dfrac{\prod_{i=1}^n \Omega(A(w^i_{mu}), R_{ju}) P(R_{ju}, I_u=1)}{\sum_{y=1}^{v_u} \bigg(\prod_{i=1}^n \Omega(A(w^i_{mu}), R_{yu}) P(R_{yu}, I_u=1)\bigg)}    
% \end{aligned}
% \end{align}
% \end{strip}

% \usepackage{graphicx}

\begin{table*}[]
\centering
\caption{5-fold CV results for the different incident detection modeling schemes. \textbf{Bold Font} indicates the schemes that have maximum values for F1 and AUC scores.}
\label{tab:performance-scores}
\resizebox{0.88\textwidth}{!}{%
\begin{tabular}{|l|l|r|r|r|r|}
\hline
\textbf{Model Scheme} & \textbf{Features} & \multicolumn{1}{c|}{\textbf{Precision}} & \multicolumn{1}{c|}{\textbf{Recall}} & \multicolumn{1}{c|}{\textbf{F1}} & \multicolumn{1}{c|}{\textbf{AUC}} \\ \hline
{[}M1, Baseline{]} Random Forest & Avg. Reliability \& Count & 31\% & 78\% & 44\% & 64\% \\ \hline
{[}M2, Baseline{]} Logistic Regression & Avg. Reliability \& Count & 34\% & 68\% & 45\% & 65\% \\ \hline
{[}M3, Proposed{]} Random Forest & Plausibility (Clustering-based) & 37\% & 64\% & 34\% & 69\% \\ \hline
{[}M4, Proposed{]} Logistic   Regression & Plausibility (Clustering-based) & 40\% & 56\% & 35\% & 69\% \\ \hline
{[}M5, Proposed{]} Random Forest & Plausibility (Segmentation-based) & 36\% & 67\% & 47\% & 65\% \\ \hline
{[}M6, Proposed{]} Logistic   Regression & Plausibility (Segmentation-based) & 36\% & 70\% & \textbf{48\%} & 66\% \\ \hline
{[}M7, Proposed{]} Random Forest & All (Clustering-based) & 30\% & 70\% & 37\% & \textbf{71\%} \\ \hline
{[}M8, Proposed{]} Logistic Regression & All (Clustering-based) & 30\% & 66\% & 36\% & \textbf{71\%} \\ \hline
{[}M9, Proposed{]} Random Forest & All (Segmentation-based) & 35\% & 72\% & 47\% & 68\% \\ \hline
{[}M10, Proposed{]} Logistic Regression & All (Segmentation-based) & 36\% & 69\% & 47\% & 68\% \\ \hline
\end{tabular}%
}
\end{table*}

\subsection{Waze Report - Incident Association}
\label{subsec:waze_incident}

Section \ref{subsec:grouping} discussed grouping of Waze reports to segments (clusters) in a time step. As we consider fusing information at discrete time steps, we will have another set of Waze reports in the next step. All the Waze reports in a segment (cluster) is assumed to be associated with an incident. 
% The Waze report - incident association is defined as follows: 

% \textit{Does each new Waze report correspond to an existing incident, i.e., an existing cluster or does it correspond to a new incident not observed in the previous time step.} 

Let $\mathbf{w}_{m+1}$ represent the set of Waze reports in time step $T_{m+1}$. Through the Waze Report - Incident Association analysis, we would like to know if a Waze report $w_{m+1}^i$ corresponds to an existing cluster $C_u$.

In order to realize the above analysis, we first generate a hypothesis that a Waze report belongs to a given cluster, and then test that hypothesis.
%\textcolor{red}{
We use process identified in Section~\ref{subsec:grouping} to associate Waze reports with one of the existing clusters or creating new ones. %}
%
%\textcolor{green}{Yasas: Please write here how you are associating Waze reports in the next step to existing clusters or creating new clusters.}

After we assign each report to a cluster, we then repeat the analysis in Section \ref{subsec:pred} to calculate the updated probabilities of incident detection and localization. It should be noted here the posterior probabilities obtained in the previous time step are used as prior probabilities in the current time step.  This use of cluster information and probabilities across two time steps is illustrated using dotted arrows in Figure \ref{fig:overview}.

\subsection{Incident Classification Model}
\label{subsec:pipeline}

After calculating the posterior probabilities for a potential incident associated with a region, we use the resulting probabilities as features into a classification model. 
%\textcolor{blue}{Yasas: this statement should come ahead of the previous statement. Our goal is to find the threshold, and we are doing that by building a classification model}
% 
In this study, we explore different types of binary classification models to identify the optimal value for the probability to create a decision boundary, specifically: %The list of classifiers we used are:
 Logistic Regression %, %Decision Tree, 
 and Random Forest.  
The classification model identifies the decision boundary to determine whether a given probability value indicates an incident or not. \textcolor{black}{The labels for each prediction time step was obtained by checking the presence of E-TRIMS accident record during the incident time interval associated with the prediction.}

%Above classifiers are set to 
We train the classification model using balanced strategy to estimate class weights since the class labels are unbalanced in our datasets. 
For training classifiers, we use \textit{L-BFGS} algorithm as optimizer for Logistic Regression and the maximum depth of 
% Decision Tree classifier and
Random Forest Classifier is set to five and three respectively. % accordingly. 
Other hyperparameters of the models are selected based on a grid search method (details in Section~\ref{sec:results}.) 

%%%%%%%%%%%%%%%%%%%%
%%%%%%%%%%%%%%%%%%%%
\section{Experimental Evaluation}\label{sec:experiments}
%\subsection{Setup} 
% - experimentation schemes for modeling --  baselines and proposed schemes \\
% \hl{please explain the date ranges and data used for the experiemnts.}
We use five-fold cross validation (CV) setup for evaluation to compare the performance of classification model schemes for incident detection. %There are no specific related works in the literature that address our problem. % in this study. 
We used the Waze and E-TRIMS data from October 2019 to December 2019 for validating our schemes. In total there were 33218 Waze reports and 2878 E-TRIMS incident records in the dataset. 
We analyze the following model schemes using varied sets of features:   
\begin{itemize}[leftmargin=*]
    \item \textbf{[M1-M2] Baselines} - Avg. Reliability + Count Features: For each time step $t_s$, this scheme uses the average score of reliability of Waze reports and the total count of Waze reports in that time step as features. %This scheme is inspired from a \textcolor{red}{prior study on Waze in transporation domain~\cite{}.}   
    
    \item \textbf{[M3-M4] Proposed} - Plausibility (Clustering-based) Features: This scheme uses our Bayesian-theoretic approach to compute the plausibility score using posterior probabilities of an incident detection in a region. The posterior inference is based on the grouping of Waze reports obtained through clustering approach (\textit{c.f.} Section~\ref{subsec:grouping}). Priors for this scheme is based on E-TRIMS data from Sept., 2019. 
    
    \item \textbf{[M5-M6] Proposed} - Plausibility (Segmentation-based) Features: This scheme is similar to the previous scheme, with one difference of computing the posterior inference for plausibility scores based on the grouping of Waze reports obtained through segmentation approach (\textit{c.f.} Section~\ref{subsec:grouping}). %. Similar to \textbf{M4-M6}, priors for this scheme is based on E-TRIMS data from September, 2019.
    
    \item \textbf{[M7-M8] All (Clustering-based)} - This scheme uses all of the above features. We used clustering approach for grouping Waze reports for this strategy. 
    
    \item \textbf{[M9-M10] All (Segmentation-based)} - Similar to previous scheme, except we use segmentation approach to group Waze reports. 
    
    %Priors Features: This scheme is based on an ideal scenario that each emergency management services could have access to the official ETRIMS type of system with historic dataset of official incident reports, which is not realistic due to access restrictions and liability issues. This modeling scheme uses the priors for incident detection in a region calculated using similar procedure in Section~\ref{sec:preprocessing}. This scheme uses all the available ETRIMS data up to the incident detection time interval and % as the predefined set of accident records. followed by computing the priors for regions, which are used as features for Logistic Regression Classifier. % to obtain the final results. 
\end{itemize}

%\subsection{Evaluation Measures} 
To evaluate the performance in 5-fold CV, we use the following metrics that are standard for classification models: Precision, Recall, F1 score, and AUC. 

\section{Result Analysis and Discussion}
\label{sec:results}

% \textcolor{green}{Yasas - Done:
% \begin{itemize}
% \item We will also need to add some description about the baseline procedures and the features they used
% \item Also, we need to discuss the performance metrics and may not need to write down the expression for them
% \item Moreover, we need to have a discussion about the various these various performance metrics (Accuracy, Precision, Recall, F1, ROC AUC) and benefits of each metric.
% \item Need to add a description of the dataset that we used for analysis: Time period, region (Nashville), number of Waze reports in that time period. 
% \end{itemize}
% \item Need to have a description of the TDOT dataset  as we are using this as ground truth
% }

This section describes the results for both incident detection models and the impact of spatio-temporal resolution in the modeling. We also present a case study to show the benefits of using crowdsourcing %based incident detection 
to complement the conventional but slower data collection method to assist emergency response. 

\textcolor{black}{After performing grid search on the hyperparameters of the proposed model, we found that hyperparameter values $T'=25~mins,~t_s=1~min,~\delta=100~meters,~Res=6$ provide the best F1 score. We used those parameters in comparing different model schemes in Section~\ref{sec:results_detection} and case study in Section~\ref{sec:results_case_study}. Detailed performance variation of model M6 with different hyperparameter configurations is provided in Section~\ref{sec:results_st_analysis}.}

% \textcolor{blue}{Yasas: Please indicate somewhere what are all the hyperparameters in this analysis. Mention that we are performing a study on the performance of the model with respect to these hyperparameters. From the results, which parameters provide the best results? When somebody else wants to perform a similar study, which parameters should be use?}

%%%%%%%%%%
\subsection{Incident Detection}
\label{sec:results_detection}
% Incident Localization Performance

% classification results using different algorithms -- Logistic Regression, Decision Tree, Random Forest. 

Table~\ref{tab:performance-scores} provides the performance of the baselines and proposed methods, where the key observations are the following. 
\begin{itemize}[leftmargin=*]
    \item  It is evident that the overall performance of the proposed modeling schemes is better than the baseline approaches in general, based on F1 score and AUC. Moreover, we observe that the results with logistic regression based schemes outperform others. 

    \item The performance of baseline classifier with Avg. Reliability and number of Waze reports as features has better recall ($78\%$) with Random Forest Classifier, however, low precision. % is low when compared with other features. 
    Consequently, the F1 score of the baseline classifier is less than the proposed approaches. 

    \item The clustering approach provided better precision at the cost of recall thereby reducing overall F1 score for schemes M3 \& M4. It is likely %that the epsilon parameter we used produced 
    due to dense but accurate clusters that resulted in better precision. 
\end{itemize}

%%%%%%%
\subsection{Spatio-Temporal Resolution Analysis}
\label{sec:results_st_analysis}

Figures~\ref{fig:incident_interval}, ~\ref{fig:aperture_size}, and~\ref{fig:time_step} 
illustrate the variation of different performance scores with respect to four Spatio-Temporal parameters in our model. 
Figure~\ref{fig:aperture_size} shows that there is a need to optimize the spatial bound on the unit of analysis for the plausibility score computation for spatial units.  
Figure~\ref{fig:incident_interval} reveals that there has been a slight increase in F1 score with increasing incident interval. %A likely explanation
It is possible that when the incident interval increases there is longer time period to consider it as an accident. 

% how the performance for incident detection varies by the change in the space and time resolution of analysis? 

\begin{figure}[ht!]
\centering
\includegraphics[width=0.45\textwidth]{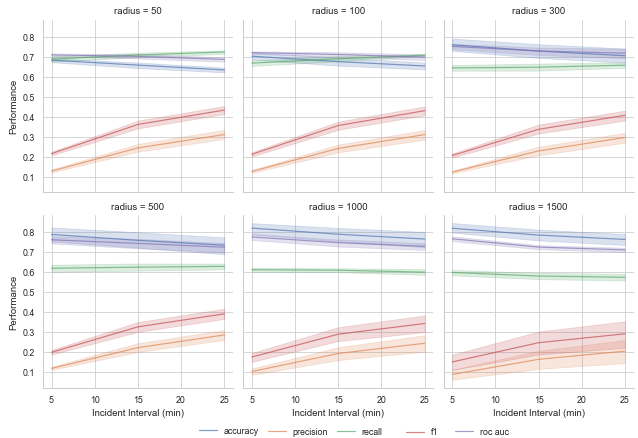}
\caption{Variation of performance with incident interval.} \label{fig:incident_interval}
\end{figure}

\begin{figure}[ht!]
\centering
\includegraphics[width=0.45\textwidth]{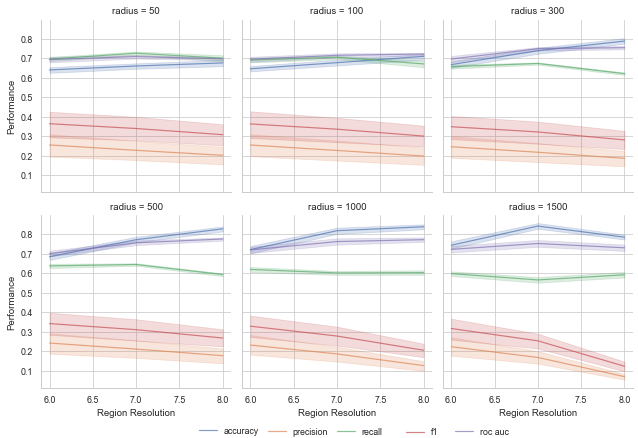}
\caption{Variation of performance with aperture size for spatial grids.} \label{fig:aperture_size}
\end{figure}

\begin{figure}[]
\centering
\includegraphics[width=0.45\textwidth]{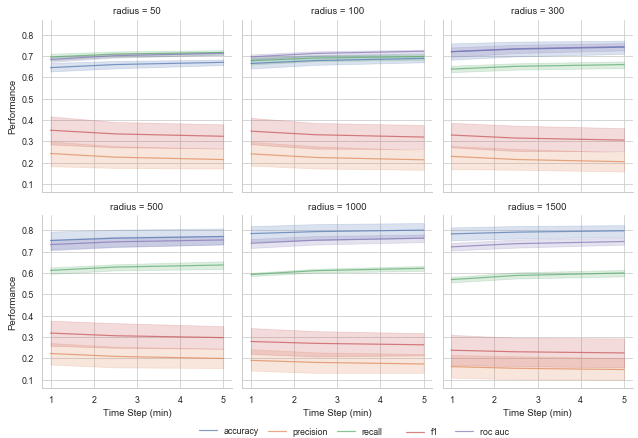}
\caption{Variation of performance with time step.}\label{fig:time_step}
\end{figure}

% How much an agency could gain the lead on response time if relied on Crowdsourced web data for alerts? 
\vspace{-0.2in}
\begin{figure}[ht!]
\centering
\includegraphics[width=0.25\textwidth]{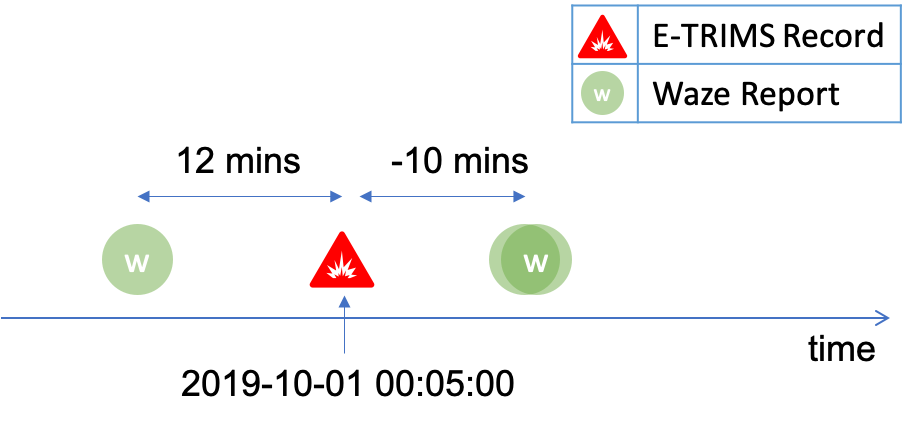}
\caption{Case study scenario for timing in accident detection in a region.} \label{fig:case_study}
\vspace{-0.2in}
\end{figure}

\subsection{Case Study} 
\label{sec:results_case_study}

This case study presents the benefits of using our approach for early detection of incidents. To exemplify, consider the incident that happened at $00:05:00$ on $2019-10-01$ (\textit{c.f.} Figure~\ref{fig:case_study}). 
Using our approach, the Waze reports 12 minutes before the E-TRIMS accident record can detect the plausibility of an incident in that region.  %Additional Waze reports after 10 minutes of recording of accident in the ETRIMS appear as well. 
We collected the time differences between the Waze reports that led to correct accident detection and the corresponding E-TRIMS records. We then averaged the detection times to get an average accident prediction time. Any prediction that comes before the official record from E-TRIMS is considered to have a positive  prediction time and a prediction that comes afterwards is considered to have a negative prediction time. We found that the average accident prediction time of our best model (Model Scheme - M6) is $5.92~minutes$. 
This evidence suggests %on average, 
our approach can help build predictive systems for incident detection using unconventional crowdsourced data, before a conventional system. % records incidents for emergency services. 

%%%%%%%%%%
\section{Conclusion}
\label{sec: conclusion}
%We demonstrate a principled approach to address the delays and limitations in rapidly collecting data about emergency incidents, through the unconventional source of crowd-generated observational reports.  
% 
This paper presents a principled methodology based on Bayesian theory to address key challenges of modeling the uncertainty and unreliability of crowd-generated reports in their integration across space and time, to detect emergency incidents earlier than the official reporting mechanisms.   
Our experiments using %four-months data from 
Waze data and official reported incidents in Nashville region validate %applicability of 
our method with relative gain in F1-score over 5\% against baselines. 
The application of this research can help emergency response operations in systematically incorporating unconventional crowdsourcing data. % sources for early incident detection.   

%\subsection{
\textit{Limitations and Future Work}. 
%While the above results demonstrate the benefits of leveraging crowdsourcing data for incident detection, 
We note the limitation in accurately matching crowdsourced data with the formal incident reports, given the human errors and systematic delays in emergency communication. We can build upon this study to explore automated feature extraction and fusion models to improve the performance, and also, improve the fidelity of the analysis (by moving from regional grids to road segments). 

%\st{reduce the unit of analysis from regional grids to the resolution at road segment levels.}  

%\textcolor{red}{
%In addition, we can explore how to learn mapping of Waze reliability to probability value based on historical data. %}
 
%\textcolor{red}{Updated:}
\spara{Reproducibility.} The code for experiments is available at \url{https://github.com/ysenarath/emergency-incident-detection-web-intelligence-2020}.

\balance

%%%%%%%%%%%%%%%%%%%%%%%%%%%%%%%%%%%%%%%%%%%%%%%%%%%%%%%%
%\input{08_acknowledgement}
\spara{Acknowledgement.} 
%The research presented in this paper was supported by funds 
We thank the U.S. National Science Foundation grants (1814958 and 1815459) and a grant from Tennessee Department of Transportation (T-DOT) for partial research support. We also thank Mr Said El Said from T-DOT, Dr. Ayan Mukhopadhyay and Dr. Sayyed Vazirizade from Vanderbilt University for data support and providing feedback.

%\st{We thank the U.S. National Science Foundation grants 1815459 and 1814958 for supporting this research. We also thank XXX for feedback and helping in data collection.}

%Authors would also like to acknowledge reviewers for valuable feedback.

%%%%%%%%%%%%%%%%%%%%%%%%%%%%%%%%%%%%%%%%%%%%%%%%%%%%%%%%

\bibliographystyle{IEEEtran}
\bibliography{wi_paper}

\end{document}